# ESR studies of ion implanted phosphorus donors near the Si-SiO$_2$ interface

Paul G. Spizzirri, Wayne D. Hutchison[†], Nikolas Stavrias, Jeffrey C. McCallum, Nakorn Suwuntanasarn[†], Libu K. Alexander[†] and Steven Prawer.

*Australian Research Council Centre of Excellence for Quantum Computer Technology*
*School of Physics, The University of Melbourne, Victoria 3010, Australia.*

[†]*Australian Research Council Centre of Excellence for Quantum Computer Technology*
*School of PEMS, The University of New South Wales, ADFA, Canberra, ACT 2600, Australia.*
[*]pgspiz@unimelb.edu.au
(Dated: February 9[th], 2010)

This work reports an ESR study of low energy, low fluence phosphorus ion implantation into silicon in order to observe the activation of phosphorus donors placed in close proximity to the Si-SiO$_2$ interface. Electrical measurements, which were used to estimate donor activation levels, reported high implant recoveries when using 14 keV phosphorus ions however, it was not possible to correlate the intensity of the hyperfine resonance signal with the electrical measurements in the presence of an SiO$_2$ interface due to donor state ionisation (i.e. compensation effects). Comparative measurements made on silicon with an H-passivated surface reported higher donor hyperfine signal levels consistent with lower surface defect densities at the interface.

## I. INTRODUCTION

Semiconductors rely upon the successful incorporation of strategically placed dopants in order to achieve specific electrical characteristics. In the current regime of nano-scale device development, fewer dopant atoms are being introduced into the active channel region bringing the assumptions of uniform (active) dopant distribution into question[1]. This question is most pertinent to the development of a solid-state quantum computer (QC) as originally proposed by Kane[2] where single phosphorus (P) donors are to be placed as arrays within the silicon host matrix with high precision. Ion implantation has been championed as a rapid development tool for the fabrication of such devices since it is a standard technique for introducing dopants into silicon and is compatible with most lithographic techniques. However, in order to reduce the ion straggle and increase positional accuracy, lower ion energies are being employed bringing dopant atoms even closer to the silicon-silicon dioxide (Si-SiO$_2$) interface which, in general, will be "imperfect".

Ion implanted devices fabricated with an SiO$_2$ layer in close proximity to the doped region (eg. MOSFET's) can also undergo dopant segregation and redistribution during processing resulting in a modification to their desired electrical characteristics or worse. For example, in the case of non-uniformly doped nano-MOSFET's[3], the threshold voltage may vary by up to 50 %. For few donor devices, this may mean the difference between having an electrically active dopant and no dopant (i.e. dopant loss). There are several reasons proposed for dopant loss near the Si-SiO$_2$ interface. For implanted structures, the dominant mechanism is thought to be associated with implantation induced point defects produced by elastically recoiling ions which create vacancy and interstitial profiles. Subsequent annealing results in two changes to the dopant concentration profile. The first, transient enhanced diffusion (TED) occurs over short time frames[4, 5] and for fluences above ~1x10$^{13}$ cm$^{-2}$. TED can cause implanted P to move towards the silicon surface[6]. This is mediated by interstitial defects and results in dopant accumulation at the Si-SiO$_2$ interface which is known to be highly efficient at trapping a significant number of P donors (up to ~2x10$^4$ cm$^{-2}$) by chemical incorporation (i.e. P compound formation) into the first monolayer of silicon[7]. Once trapped at the interface, impurity atoms are inactive and no longer contribute to the electrical properties of the device[8]. As a result of this mechanism, the interface can play a crucial role in the formation of shallow junctions when using low energy implanted ions. In addition, the gradients driving this redistribution may result in higher local impurity concentrations (i.e. clustering) just below the interface depending upon the implantation profile and diffusion kinetics. The second annealing process, which occurs at longer time scales, is in-diffusion. This results in P redistribution away from the interface. These mechanisms are not well understood yet they are clearly dependent upon the implant energy, fluence and annealing parameters employed[6]. To date, there are no reports on the TED of 14 keV P$^+$ implants at fluences <10$^{13}$ cm$^{-2}$.

Single ion placement has been successfully demonstrated for P in silicon[9] in QC architectures however, it is not known whether (i) the silicon lattice is fully recovered and (ii) donor atoms are successfully incorporated following standard thermal processing since in this device geometry, there are no redundant donor atoms. Questions about the success of impurity incorporation have been raised by other workers who have observed particularly poor activation levels for low energy implanted donors placed in close proximity to an interface as determined by the spreading resistance method[10]. In this work we report an ESR study of shallow, low implantation energy, low fluence P doping in silicon and discuss the impact of the Si-SiO$_2$ interface on donor spectroscopy. Electron spin resonance is also well suited to the study of the Si-SiO$_2$ interface as the defects of interest (e.g. P$_b$ centres) were originally observed and characterised by Nishi using this technique[11, 12]. The intensity of the P hyperfine resonance signal is therefore considered as a function of the interface quality and donor proximity to the surface.

## II. EXPERIMENT

**A. Sample preparation**

High resistivity silicon ($\rho$ > 4 k$\Omega$.cm) was cleaned at each stage of the process using standard piranha and RCA2 protocols using ULSI grade reagents in a cleanroom. Silicon ion (Si$^-$) implants were performed using the 150 keV ion implanter located at the Australian National University whereas phosphorus ions (P$^+$) originated from the Melbourne Colutron low energy ion implanter. Room temperature, two dimensional (areal) implants were obtained by irradiation through a metal aperture using an electrostatic deflection system to obtain uniform dosimetry. Phosphorus ion fluences ranging from 10$^{11}$ up to 10$^{13}$ P cm$^{-2}$ were produced at various energies without tilt or rotation for both atomic and molecular ion beams. All phosphorus implantations were carried out along the [001] axis. Silicon implantation was performed with samples tilted and rotated at 10°. Post-implantation annealing was performed at 1000 °C for 5 seconds in an argon atmosphere using a rapid thermal annealer (ModularPro RTP) to electrically activate the implanted impurities and repair the host silicon lattice. It is anticipated that P impurities processed in this way will be substitutionally incorporated into the silicon host lattice. Controlled silicon native oxide surfaces were grown in a cleanroom environment after initial sample cleaning and oxide removal. Thicker oxides (5 nm) were grown at 820 °C in a dry oxidation furnace. Samples with surface oxides were only cleaned immediately prior to implantation and annealing. Surface hydrogen termination was achieved by treatment with hydrofluoric acid. This was performed immediately before implantation, annealing and ESR measurement to ensure that the surface was oxide free. H-passivated surfaces were stored in degassed propan-2-ol to slow surface re-oxidation between processing. Sheet resistance ($R_s$) values were obtained at room temperature by four point probe measurement.

**B. CW-ESR measurements**

Electron spin resonance (ESR) measurements were performed on a Bruker ESP300 spectrometer with a standard X band cavity. This was coupled to a flow cryostat controlled by an Oxford Instruments ESR-900 cryogenic sample temperature controller. Since the bound donor (Si:$^{31}$P) spin lattice ($T_1$) relaxation time is reported to be very long (>$10^3$ s) at 2 K but strongly dependent on temperature (varying by 7 orders of magnitude between 3 and 20 K)[13], measurements were mostly performed at higher temperatures (14 K) to ensure a reasonable measurement time through the shortening of $T_1$.[14]

Samples prepared for this study were in the form of bars (4.9 mm x 10 mm x 0.5 mm) cleaved from silicon wafers with a (001) surface orientation and mounted on a high-purity silica-glass rod for insertion into the microwave cavity. Care had to be exercised during the handling of the samples to prevent the formation of additional defects on the bar surfaces or edges. This was achieved through the use of Teflon tooling by contacting the bar edges only during handling with residual dominant defects attributed to sample scribing and cleaving. Field sweeps of width 100 G about a centre field of 3380 G were employed along with a microwave frequency of ~9.45 GHz. Measurements were performed either in the dark (*dark*) or under white light illumination (*light*) using a halogen lamp to further shorten $T_1$ through the interaction of donor spins with photoelectrons. This also permitted ESR measurement under *non-equilibrium* conditions (i.e. optical excitation) which is known to give rise to an ESR donor signal enhancement through a mechanism which thought to involve electron capture from photo-generated electron-hole pairs (e-h) by ionized donor states[15]. The spectrometer displayed sensitivity to ~5x10$^{10}$ spins and resonance lines have been fitted with differential Lorentzians and thus identified according to their g-factors.

**C. Modelling**

As-implanted dopant distributions were modeled using the SRIM package[16]. One dimensional process modeling was performed using the 1D Suprem-IV numerical simulator[17]. As implanted P atom distributions created by SRIM were imported into Suprem-IV to model dopant diffusion (using the standard model). Molecular P implants were modeled assuming each ion has half the implant energy of the implanted molecule. This approach has been compared with molecular dynamics simulations and shown to represent low energy molecular implantation processes well[18].

## III. RESULTS

The one-dimensional, simulated substrate doping profiles for each of the annealed P implants studied are shown in Figs. 1(a) to (d). The implanted fluence, surface termination and estimated loss (%) to the surface oxide are all indicated. Channelling of the ion beam into hydrogen terminated silicon is also anticipated and modelled in (d). The simulations show the as-implanted dopant distribution added to a nominal background doping of the substrate (assumed to be around $10^{14}$ cm$^{-3}$). Following thermal processing (RTA), the P is redistributed using a standard diffusion model and depicted in the "annealed" plots. They show that the dopant diffusion range is expected to be small under these processing conditions which is important for nano-devices.

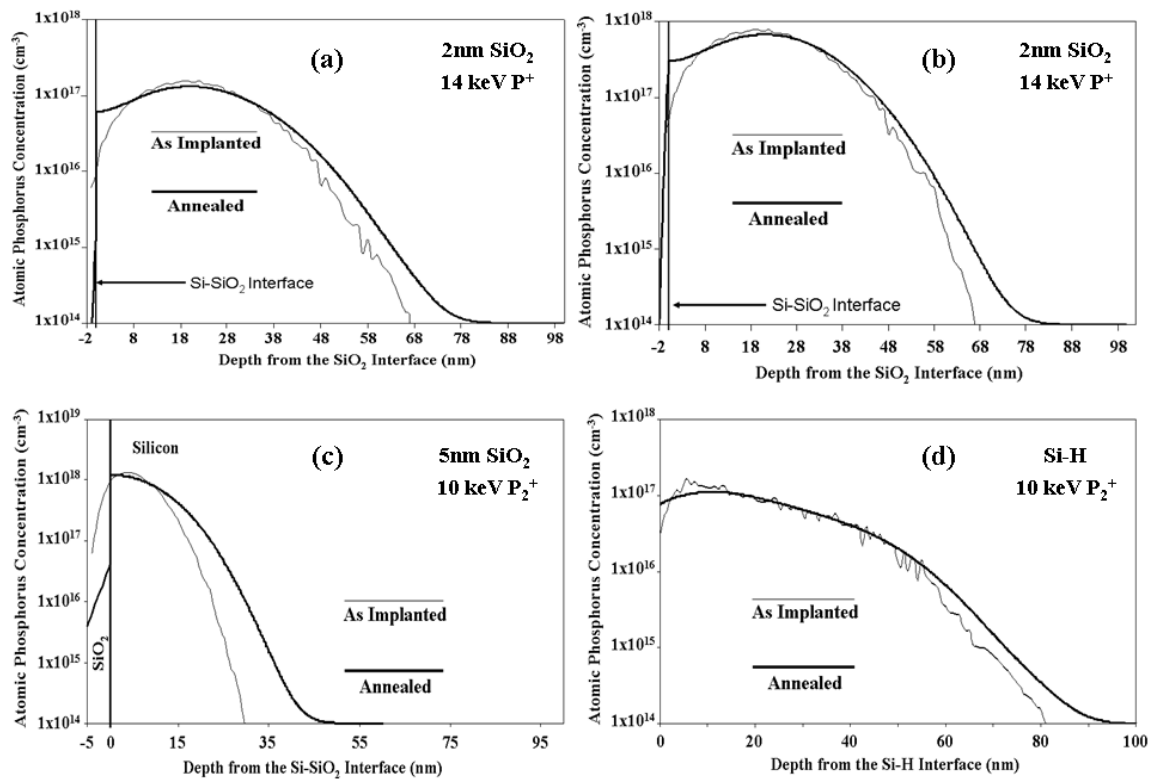

FIG. 1. Process simulations for the phosphorus ion implanted samples studied in this work. (A) 14 keV atomic P$^+$ ions implanted (fluence = 4.1x10$^{11}$ cm$^{-2}$) through a 2 nm surface oxide, (B) 14 keV atomic P$^+$ ions implanted (fluence = 2.1x10$^{12}$ cm$^{-2}$) through a 2 nm surface oxide, (C) 10 keV molecular P$_2^+$ ions implanted (fluence = 8.0x10$^{11}$ cm$^{-2}$) through a 5 nm surface oxide and (D) 10 keV molecular P$_2^+$ ions implanted (fluence = 2.0x10$^{11}$ cm$^{-2}$) through a H-passivated surface. The peak atomic P concentration has not been adjusted to correct for loss in the surface oxide.

The sensitivity of the ESR technique to the measurement of paramagnetic centres makes it an ideal tool for monitoring implantation and annealing processes to determine whether ion cascade induced defects result in the creation of stable paramagnetic silicon defects. For example, vacancy point defects (and their clusters) can assume positive, neutral and negative charge states which are identified as $V^+$, $V^0$ and $V^-$ with the $V^+$ and $V^-$ states paramagnetic[19]. If the thermal repair of this damage is incomplete, electrically active point defects remain. Silicon ions are able to produce point, cluster and extended defects in silicon[20] and have a similar damage profile to that of P ions but don't introduce dopant states making it a good choice for characterizing the effectiveness of the implantation and thermal processes used in this work. Figure 2 shows ESR measurements of high resistivity silicon which has been (a) untreated and (b) implanted with 14 keV $Si^-$ ions (fluence = $1 \times 10^{12}$ $cm^{-2}$) followed by RTA processing. In both measurements, a broad feature dominates the spectrum at 3364.3 G (g = 2.0055) corresponding to $P_b$ ("dangling bond") defects at the $Si-SiO_2$ interface[11, 12]. The fitted peak area of the $P_b$ resonance for the implanted sample (b) is approximately double that of the untreated sample in (a). An increase in the $P_b$ contribution to the spectrum would likely be associated with an increase in the density of interfacial (electrical) trap states ($D_{its}$). This would be consistent with the work of Peterström[21] on unpassivated, phosphorus implanted (50 keV) oxides where $D_{its}$ values were found to be sensitive to the ion implantation fluence, increasing by an order of magnitude over the range $5 \times 10^{11}$ to $4 \times 10^{12}$ P $cm^{-2}$.

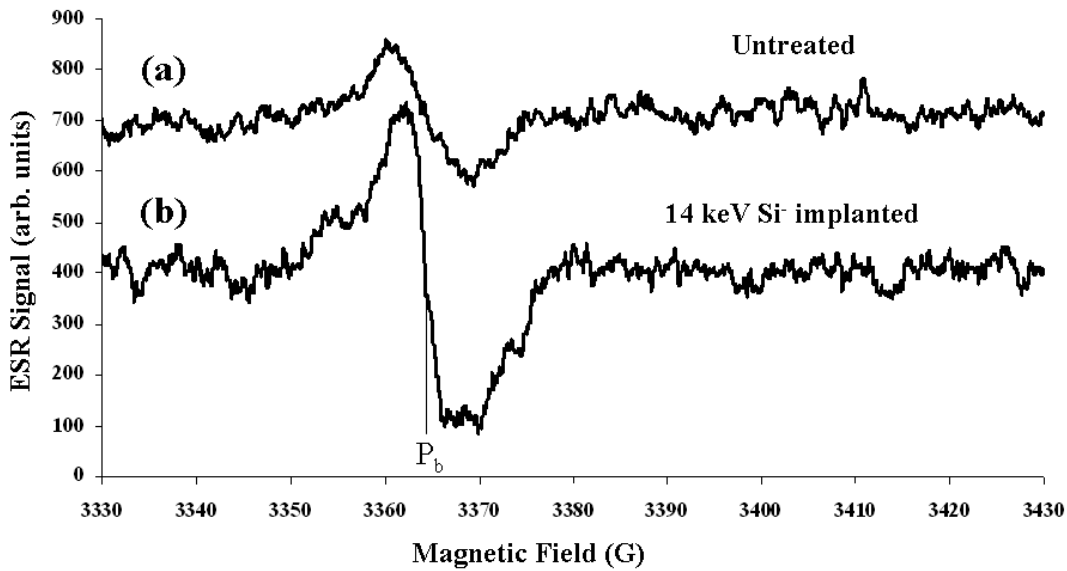

FIG. 2. *Dark* ESR measurements (14 K) of high resistivity silicon ($\rho > 4$ k$\Omega$.cm) with a 2 nm native surface oxide (a) measured as supplied (untreated) and (b) implanted with 14 keV $Si^-$ ions to a fluence of $1 \times 10^{12}$ $cm^{-2}$ followed by thermal annealing. The $P_b$ label with marker identifies the approximate location of the $P_b$ defect resonance feature. Plots have been offset.

The $P_b$ states are known to be one of the most technologically important defects for MOS devices comprising trivalent silicon centres ($\bullet Si \equiv Si_3$)[22]. It should be noted that measurements in this work include signals obtained from all surfaces of the sample, not just the "top" implanted surface. This means that some of the interface signal originates from damage introduced during substrate preparation and handling (e.g. scribing and cleaving). While it is clear that there is an increase in the $P_b$ signal intensity associated with the silicon implant (compared to the untreated sample), no other paramagnetic centres are evident. This suggests that the processing conditions adopted in this work are sufficient to remove all but the interface related defects. These interface defects are problematic for nano-device applications however, it should be noted that surface oxides are frequently removed and regrown or repaired by annealing in a hydrogen-containing ambient to reduce the density of electrically active trap states. This process is thought to involve cracking centres which produce atomic hydrogen capable of passivating the dangling bond defects such as $P_b$ centres and E' defects[23].

While ESR can be used to study P in silicon as electrically active (shallow) donor states[24], it can also be used to observe P related paramagnetic defects such as the P vacancy pair in silicon (i.e. E-centre)[25]. Spectral identification of donor states is based upon the observation of the hyperfine interaction of the donor electron with the magnetic moment of the $^{31}$P nucleus (I = ½, 100% natural abundance) giving rise to a spectral doublet (~42 G splitting with g = 1.99875±0.00010)[23]. It is known that for P concentrations below $10^{16}$ cm$^{-3}$, ESR measurements report only 2 hyperfine split lines arising from isolated donors[26]. For P concentrations around $10^{17}$ cm$^{-3}$, evidence of exchange coupling is expected as a central line feature (g = 1.99875). At concentrations (i) approaching and (ii) exceeding the metal-insulator (M-I) transition (~3.7x10$^{18}$ cm$^{-3}$)[27], the hyperfine signal is replaced by (i) a dopant impurity band (DIB) (g = 1.99875) and finally, (ii) a free (conduction band) electron line (g = 1.9995)[28].

Figure 3 shows measurements from several 14 keV P$^+$ implants prepared with fluences ranging from 4x10$^{11}$ cm$^{-2}$ up to 1x10$^{13}$ cm$^{-2}$ at which fluence, metallic donor states should form (i.e. [P]$_{peak}$ = 5x10$^{18}$ cm$^{-3}$). Sheet resistance values ($R_s$) are reported for preparations (a) and (b) which confirm high levels of dopant activation consistent with the annealed (simulated) dopant profiles. Substantial dopant segregation at the interface (i.e. P pileup) is not evident in these measurements as this would render P donors electrically inactive resulting in higher sheet resistance values as has been reported previously[8]. The level of donor activation (electrical) obtained in this work was estimated by comparing the measured $R_s$ values with those calculated using the following empirical expression[29]

$$R_s = KF^{-0.7}\left(1.56\times10^{-6}(LnE)^2 - 9.5\times10^{-6}(LnE) + 1.66\times10^{-5}\right)^{-0.3} \quad (1)$$

where $K$ is a constant with a value of $5.8\times10^{10}$ ohm.cm$^{-1.1}$ for donors, $F$ is the implant fluence (ions/cm$^2$) and $E$ is the implant energy (keV). The activation values (%) for the two lowest fluence sample preparations, which demonstrate high implanted donor fluence recoveries, are reported in Fig. 3.

The data in Fig. 3 shows clear signs of $P_b$ defects, isolated and exchange coupled P donors and metallic (free electron) states in spectrum (c). These results are consistent with the process simulations of Fig. 1 if we assume that the implanted ions are mostly electrically active, as suggested by the measured values of $R_s$, with distributions described by the annealed profiles. Comparing measurements in (a) and (b), we observe very similar donor resonance profiles with equivalent signal intensities along with weak exchange (central) resonance features which also have similar fitted areas. It is clear from these measurements that the donor hyperfine resonance signal intensity does not reflect the order of magnitude difference in the implanted fluence between these samples. In addition, the exchange resonance (central) feature of spectrum (b) is considerably smaller than expected given the peak donor concentrations predicted in the annealed profile of Fig. 1.

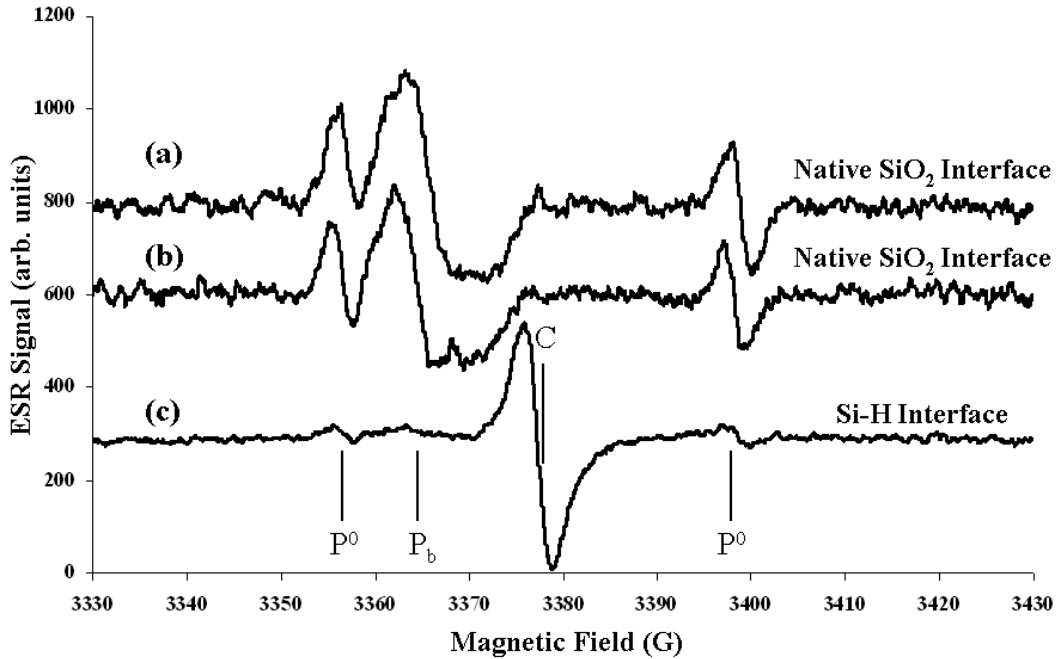

FIG. 3. ESR measurements of 14 keV $P^+$ implants into silicon with native oxide and H-passivated surfaces. Samples were implanted to the following nominal fluences: (a) *dark* at 9 K, fluence = $2.1 \times 10^{12}$ P cm$^{-2}$ ($R_s$ = 9000 ohm/□, ~70% activation), (b) *dark* at 14 K, fluence = $4.2 \times 10^{11}$ P cm$^{-2}$ ($R_s$ = 36,000 ohm/□, ~50% activation) and (c) *light* at 5 K, fluence = $1.0 \times 10^{13}$ P cm$^{-2}$ (peak P concentration $[P]_{peak}$ = $5 \times 10^{18}$ cm$^{-3}$). The $P^0$ and C labels with markers identify the approximate locations of isolated and exchange coupled (metallic) P donor resonance features. Surface preparations are indicated and plots have been offset.

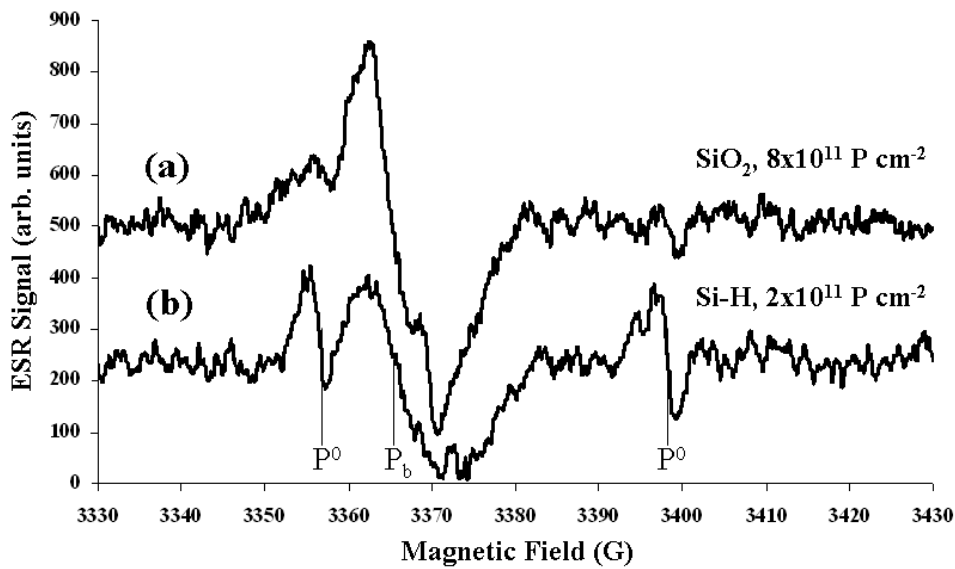

FIG. 4. *Light* ESR measurements taken at 14 K from (a) 10 keV $P_2^+$ ions implanted through a 5 nm SiO$_2$ layer to a fluence of $8 \times 10^{11}$ cm$^{-2}$ and (b) 10 keV $P_2^+$ ions implanted through an H-passivated monolayer to a fluence of $2 \times 10^{11}$ P cm$^{-2}$. Plots have been offset.

The third spectrum reported in Fig. 3(c) results from an implanted fluence ($[P]_{peak} = 5\times10^{18}$ cm$^{-3}$) greater than the M-I transition into a substrate with an H-passivated surface. This measurement shows a strong central line feature from a combination of exchange coupled (including clustered, hyperfine collapsed) donors, conduction band states and a smaller doublet arising from uncoupled and exchange coupled donors. The absence of an SiO$_2$ surface on this sample also results in a substantially reduced P$_b$ defect resonance signal in this spectrum.

The measurements reported in Fig. 4 arise from two samples which have been implanted with 10 keV molecular P$_2^+$ ions through (a) a 5 nm SiO$_2$ layer and (b) a H-passivated surface. Once again, the broad feature dominating the spectrum at 3365 G arises from P$_b$ centres and is more intense for the sample with an SiO$_2$ interface as expected. In addition to this, a spectral doublet (i.e. hyperfine satellites) with ~42 G splitting centered at ~3377.5 G (g = 1.9987) is also evident and can be attributed to P donors. The measurements of Fig. 4 contrast the interface preparations and their impact on the donor resonance signal intensity. Both samples were implanted with 10 keV P$_2^+$ ions at low fluences resulting in dopant profiles that are even closer to the interface than the measurements of Fig. 3. The first preparation (a) has a 5 nm surface oxide while the second (b) is H-passivated. While both spectra show evidence of P$_b$ defect states with similar areas, the P resonance intensity is stronger in the presence of the H-passivated surface (~4 times the fitted area) even though the implanted fluence is four times lower than that of the oxide terminated surface. Ion losses to the oxide are not expected to exceed ~15 % and therefore, cannot account for the reduced signal intensity of this sample. In addition, these measurements were performed under white light illumination, the consequences of which are discussed below.

## IV. DISCUSSION

Evaluating donor states near the Si-SiO$_2$ interface presents a challenge for the ESR technique since the signal intensity is known to be dependent upon a number of parameters including: the spin relaxation time (T$_1$), frequency and amplitude of the modulating field, degree of dopant activation, dopant density and the intensity of the microwave field. Furthermore, the dopant signal is affected by donor ionisation since only the neutral charge, ground state of P donors (P$^0$) contribute a resonance signal to the ESR measurement (i.e. paramagnetic states)[30]. Donor ionization pathways include: hole capture from the valence band[31] and electron emission to the conduction band (P$^0 \rightarrow$ P$^+$ + e), a process that can be driven thermally given the shallow nature of this impurity state in silicon. Carriers which have been emitted to the conduction band are subsequently captured and retained by other centres (e.g. deep trap states) or may undergo recombination via a two step process whereby electrons and holes are captured successively as described by the Shockley-Read-Hall model[32,33,34].

From the work of Lenahan *et al* [35], it is known that unoxidised silicon atoms at the Si-SiO$_2$ interface give rise to silicon P$_b$ and silica E' defects. The P$_b$ centres establish themselves within silicon's band gap where they can trap or deplete charge carriers and so effect the position of the Fermi energy. Specifically, P$_b$ centres are known to be amphoteric; positively charged and diamagnetic when the Fermi level is near the valence band and negatively charged and diamagnetic when the Fermi level is near the conduction band. When the Fermi level is mid gap, $P_b$ centres are paramagnetic (i.e. ESR active) and have no net charge. It has been reported that the generation of $P_b$ defects accompanies the interface trap formation process and that there is a ~1:1 correspondence between the density of $P_b$ centres and density of interface traps created. It has also been suggested that if the dopant levels are high enough, a number of doped silicon layers adjacent to the interface will be depleted of carriers by these traps[36].

Systems with efficient trap states (i.e. high capture yields) can localise carriers for a sufficiently long time at low temperatures to establish an appreciable steady-state carrier occupation[37]. This could result in a persistent, ionized donor state ($P^+$) population. It has also been suggested that P donors can have their electrons transferred directly to deep centres without going via conduction or valence band states. This has been reported for the deep vacancy oxygen complex (V-O) where carriers subsequently underwent recombination at the V-O complex[38]. The efficiency for charge transfer can be very high, particularly if the defect is located within tunneling range of the dopant. Given the mid-gap energy levels of $P_b$ related defects[35], this type of compensation mechanism may well be responsible for the reduced signal levels observed for the samples prepared with defective oxide interfaces as reported in Figs. 3(a) and (b).

These carrier trapping scenarios may play a role in *static* ESR measurements with many of the trap states populated during sample cool down. By contrast, *non-equilibrium* ESR measurement conditions, established using continuous white light illumination with > $E_g$ excitation, result in the creation of large numbers of electron-hole (e-h) pairs. From the van Roosbroeck-Shockley relation[39], there are equal rates of particle creation and annihilation, with the latter occurring via both radiative and non-radiative pathways. Of the non-radiative pathways, excitonic Auger capture is thought to be one of the most important mechanisms for recombination in silicon and occurs when a free exciton meets an impurity or defect state. For an ionised donor, the electron from the exciton is captured ($P^+ + e \rightarrow P^0$) with the excess energy transferred to the hole leaving it highly excited and ejected into the valence band. The capture coefficients for this process, which have been reported as being relatively insensitive to the energy of the defect (or trap state), are significant in the Si:$^{31}$P system ($10^{-6}$–$10^{-7}$ cm$^3$ s$^{-1}$)[34]. Alternatively, free electrons and holes (in the form of e-h pairs) may be trapped directly by ionised donors (e.g. $P^+$) and acceptors (e.g. $B^-$) in compensated, bulk doped silicon[40]. This process is associated with donor signal enhancement during ESR measurement[15] and passage effects have been ruled out of the mechanism. In this enhancement scheme, the rate of e-h pair generation must be greater than the rate of electron transfer from $P^0$ to $B^0$ to allow a build-up of the $P^0$ steady-state concentration. Several of these pathways are shown schematically in Fig. 5.

The enhancement mechanisms involving the Auger-like process discussed above were clearly not observed in Fig. 4(a) which was taken from a sample with a surface oxide and $P_b$ defects. This *non-equilibrium* measurement displayed a very weak P hyperfine resonance signal and no central line feature even though the implanted fluence was four times higher than the interface oxide-free companion sample in (b). This suggests that there is a significant near surface e-h pair recombination pathway involving defect states associated with the surface oxide which efficiently competes with $P^+$ (ionized phosphorus donors) for electron capture.

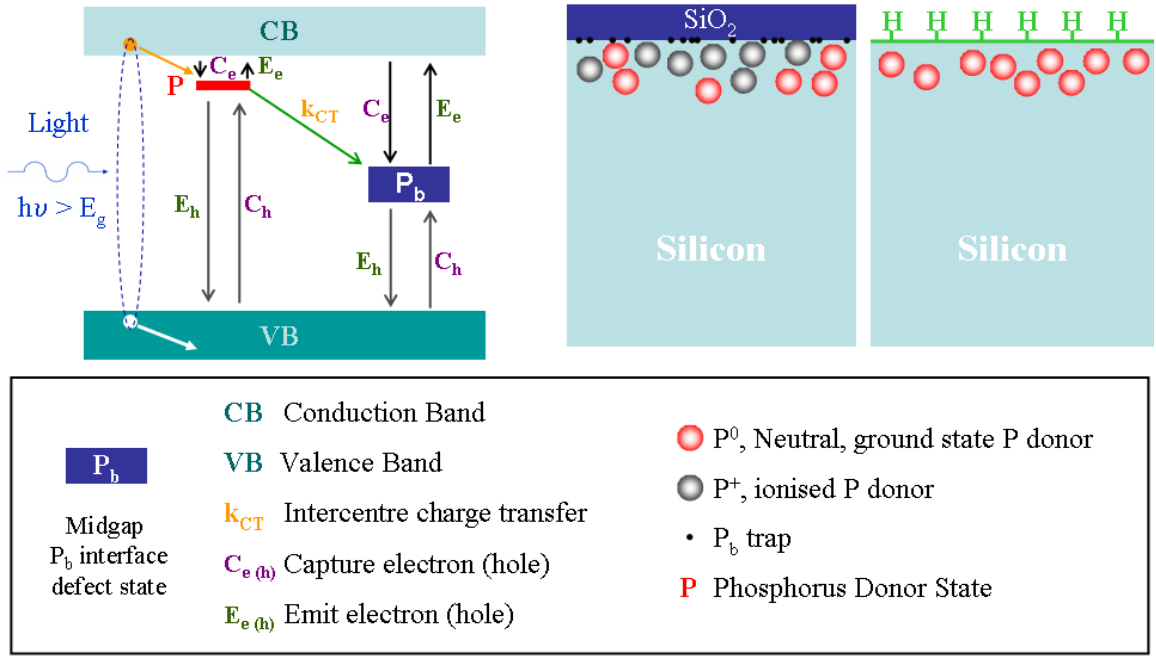

FIG. 5. Schematic band diagram of silicon showing several carrier emission, capture and recombination pathways at donor and defect sites. A direct inter-centre charge transfer process is also depicted along with a stylized interface model depicting P donor compensation.

The oxide companion sample shown in Fig. 4(b) has a H-passivated surface and displays significant P resonance signal intensity, especially given the very low implant energy and fluence employed. The Si-H termination is often described as an ideal surface providing an atomically flat and defect free (i.e. low disorder) interface with surface state densities reported[41] to be below $10^{10}$ cm$^{-2}$. It comprises a hydrogen monolayer attached to oxide free silicon and is considered a good candidate for atomic scale electronic devices where electrons are to be coupled to single atoms[42]. This measurement shows that by removing the defective states associated with the SiO$_2$ interface, a greater ESR P hyperfine signal to noise ratio can be obtained.

## V. CONCLUSIONS

In this work, we have demonstrated the utility of CW-ESR as a tool for characterising P ion implanted silicon where dopants are placed in close proximity (<50 nm) to a silicon interface. Low energy, low fluence P implants were studied and shown to be electrically active, even when using ion energies as low as 5 keV and fluences around $10^{11}$ cm$^{-2}$. The implantation/annealing process did not result in the introduction of new paramagnetic centres in the implanted region. Electrical measurements were used to estimate the levels of donor activation which were found to be very high ($\geq$ 50 %) for the 14 keV implants which is important for QC applications with single or few donor atoms. There was no evidence of transient enhanced diffusion or interfacial segregation for any of the 14 keV implants as these dopant redistribution processes would have influenced both the electrical and spectroscopic (ESR) measurements by changing the local donor state densities.

Surprisingly, the ESR P hyperfine signal intensity could not be used to quantitatively assess the degree of dopant activation in these preparations. The interfacial Si-SiO$_2$ defects, which can act as electrical traps, were found to cause the ionisation of donor states as evidenced by a reduction in the intensity of the resonance signal in *static* ESR measurements. *Non-equilibrium* measurements also exhibited lower P hyperfine resonance signal levels when compared with companion samples prepared with an H-passivated surface. This suggests that the SiO$_2$ interface defect states may be responsible for either enhanced e-h pair recombination or efficient inter-centre charge transfer.

## ACKNOWLEDGEMENTS

This work was supported in part by the National Security Agency (NSA) under Army Research Office (ARO) contract number W911NF-08-1-0527 and also by the Australian Research Council (ARC). Electronic materials engineering at the Australian National University is gratefully acknowledged for providing access to their implantation facilities.